# TESS HABITABLE ZONE STAR CATALOG

L. Kaltenegger[1], J. Pepper[2], K. Stassun[3], R. Oelkers[3]

[1] Carl Sagan Institute, Cornell University, Space Science Institute 312, 14850 Ithaca, NY, USA,
lkaltenegger@astro.cornell.edu, Tel: +1-607-255-3507
[2] Lehigh University, Physics Department, Bethlehem, PA 18015, USA
[3] Vanderbilt University, Physics & Astronomy Department, Nashville, TN 37235, USA

ABSTRACT

We present the Transiting Exoplanet Survey Satellite (TESS) Habitable Zone Stars Catalog, a list of 1822 nearby stars with a TESS magnitude brighter than $T=12$ and reliable distances from Gaia DR2, around which the NASA's TESS mission can detect transiting planets, which receive Earth-like irradiation. For all those stars TESS is sensitive down to 2 Earth radii transiting planets during one transit. For 408 stars TESS can detect such planets down to 1 Earth size during one transit. For 1690 stars, TESS has the sensitivity to detect planets down to 1.6 times Earth-size, a commonly used limit for rocky planets in the literature, receiving Earth-analog irradiation. We select stars from the TESS Candidate Target List, based on TESS Input Catalog Version 7. We update their distances using Gaia Data Release 2, and determine whether the stars will be observed for long enough during the 2 year prime mission to probe their Earth equivalent orbital distance for transiting planets. We discuss the subset of 227 stars for which TESS can probe the full extent of the Habitable Zone, the full region around a star out to about a Mars-equivalent orbit. Observing the TESS Habitable Zone Catalog Stars will also give us deeper insight into the occurrence rate of planets, out to Earth-analog irradiation as well as in the Habitable Zone, especially around cool stars. We present the stars by decreasing angular separation of the 1AU equivalent distance to provide insights into which stars to prioritize for ground-based follow-up observations with upcoming extremely large telescopes.

1. INTRODUCTION

Several thousand extrasolar planets have already been detected orbiting a variety of host stars and provide a first glimpse of the diversity of other worlds (e.g., reviewed in Udry & Santos 2017, Winn & Fabrycky 2015, Kaltenegger 2017). Among them, the first dozens of potentially habitable planets have been identified (see e.g. review by Kaltenegger 2017).

Previous ground-based wide-field transit surveys like WASP, TrES, HATNet, KELT, XO, QES (Bakos et al. 2004, Alonso et al. 2004, McCullough et al. 2005, Pollacco et al. 2006, Pepper et al. 2007, Alsubai et al 2013) have been able to detect mostly gaseous planets, while dedicated space-based surveys like CoRoT (Auvergne et al. 2009) and Kepler (Borucki et al 2010) have been designed to only survey a small region of the sky with high precision.

In contrast, the new Transiting Exoplanet Survey Satellite (TESS) is designed to search the whole sky for extrasolar planets and also to search for potentially habitable exoplanets around the closest and brightest stars. TESS is a NASA Explorer space mission with four wide-field cameras, and combines the advantage of wide-field surveys with the photometric precision and long intervals of uninterrupted observation from space. However, in order to complete the survey within the primary mission time of two years, TESS will monitor stars for 27 days near the ecliptic, with overlapping sectors yielding observational time baselines of up to almost a year near the ecliptic pole, in a region that overlaps with the JWST continuous observing zone (Ricker et al. 2016). TESS monitors several hundred thousand Sun-like and smaller stars for transiting planets and prioritizes bright stars, which allow detailed follow-up measurements of properties including planetary mass and atmospheric composition. TESS has published its first detected planets (see e.g. Huang et al 2019, Vanderspeck et al 2019).

For which of the 200,000 to 400,000 stars selected for exoplanet observations in the TESS Input Target Catalog could TESS detect planets like our own? Here we present the TESS Habitable Zone Star Catalog: Our catalog lists the stars for which TESS can detect planets, which receive the equivalent irradiation as Earth. For a subset of these stars TESS can detect planets in the whole Habitable Zone (HZ). We use the TESS Input Catalog (TIC; Stassun et al 2018) as the basis for our analysis, and update the distances with Gaia Data Release 2 (DR2; Gaia Collaboration et al. 2018).

What distinguishes this catalog from previous work like HabCat (Turnbull & Tartar 2003), DASSC (Kaltenegger et al 2010) and CELESTA (Chandler, McDonald, & Kane 2016) is that the stars included here are specifically selected to have sufficient observation time by TESS for transit detection out to the Earth-equivalent orbital distance. We also use Gaia DR2 data, which allows us to exclude giant stars from the star sample and provides reliable distances for our full star sample. All the stars have been included in the TESS exoplanet Candidate Target List, ensuring that they will also have 2-minute cadence observation (provided they do not fall in TESS camera pixel gaps), providing a specific catalog for the TESS mission of stars where planets in the Habitable Zone can be detected by TESS. This data will be available to the community in the ongoing public TESS data releases.

Several studies have explored the occurrence rate of planets from the known planets around different host stars (see recent review in Kaltenegger 2017, Savranksy et al 2018) and reach different conclusions. Simulations for the projected TESS planet yield have been discussed in detail in previous papers (Barclay et al. 2018, Bouma et al 2017, Sullivan et al. 2015, Huang et al 2018) based on a range of assumptions for planet occurance rate (see section 4.3). Note that we do not make assumptions about the planet occurrence rate here, and instead focus on the stars around which TESS can detect 2 transits for a planet receiving Earth-analog irradiation in its prime mission observation time. Observing the TESS Habitable Zone Catalog Stars will give us deeper insight into the occurrence rate of planets out to Earth-analog irradiation as well as in the Habitable Zone, informing the occurance rate discussion for planets around cool star.

Section 2 describes the Habitable Zone concept we use in our analysis, how we derived our catalog from the TIC-7, and how we updated the stellar distances using Gaia DR2, how we eliminated giant stars from the catalog and how we model detectability of transiting planets with different sizes. Section 3 presents the results, and explores for how many stars TESS can probe the whole HZ, as well as for how many stars TESS can detect planets receiving Earth irradiation as a function of their size and which are in the James Webb Space telescope continuous and extended viewing zone. Section 4 discusses our results.

2. METHODS

We first identify the stars for which TESS can detect planets out to the orbital region with similar top of the atmosphere (TOA) flux as Earth receives, using the luminosity of the star to calculate the 1 AU equivalent orbital distance from the star. This calculation is independent of models of a planet.

2.1. HABITABLE ZONE LIMITS USED

We then analyze for how many stars TESS can observe 2 transits of planets with orbital periods that place them within the HZ of their star. The HZ is a tool that guides missions and surveys in prioritizing planets for time-intensive follow-up observations.

The HZ is defined as the region around one or multiple stars in which liquid water could be stable on a rocky planet's surface (e.g., Kasting et al. 1993, Kaltenegger & Haghighipour 2013, Kane & Hinkel 2013), facilitating the detection of possible atmospheric biosignatures. Although planets located outside the HZ are not excluded from hosting life, detecting biosignatures remotely on such planets should be extremely difficult. Liquid surface water is used because it remains to be demonstrated whether subsurface biospheres, for example, under an ice layer on a frozen planet, can modify a planet's atmosphere in ways that can be detected remotely.

The width and orbital distance of a given HZ depends to a first approximation on two main parameters: incident stellar flux and planetary atmospheric composition. The incident stellar flux depends on stellar luminosity, stellar spectral energy distribution, stellar spectral energy distribution, eccentricity of the planetary orbit, and the planet's orbital distance (semimajor axis). The warming due to atmospheric composition depends on the planet's atmospheric makeup, energy distribution, and resulting albedo and greenhouse warming. In the literature, very different values of stellar irradiance are used as boundaries for the HZ. Here we use the empirical habitable zone boundaries, which were originally defined using a 1D climate model by Kasting et al. (1993), and updated in (Kopparapu et al. 2013, Ramirez & Kaltenegger 2016, Ramirez & Kaltenegger 2017), for main-sequence (MS) stars with effective temperatures ($T_{eff}$) between 2,600 K and 10,000 K. Table 1 provides values to estimate the size of the HZ, and also includes HZ limits based on 3D atmospheric models for our Sun (Leconte et al. 2013) for different host stars (Ramirez & Kaltenegger 2014) for comparison.

Equation 1 gives a third-order polynomial curve fit of the modeling results for A- to M-type host stars as shown in Kaltenegger (2017) based on values derived from models by (Kasting et al 1993, Kopparapu et al. 2013, 2014 and an extension of that work to 10,000K by Ramirez & Kaltenegger 2016):

$$S_{eff} = S_{sun} + a \cdot T^* + b \cdot T^{*2} + c \cdot T^{*3} \quad (1)$$

where $T^* = (T_{eff} - 5780)$ and $S_{sun}$ is the stellar incident values at the HZ boundaries in our solar system. Table 1 shows the constants to derive the stellar flux at the HZ limits valid for $T_{eff}$ between 2600K – 10000K: The inner boundaries of the empirical HZ (Recent Venus, RV) as well as an alternative inner edge limit modeled by 3D Global Climate model (3D; Leconte et al.2013) and the outer limits (Early Mars, EM). Note that the inner limit of the empirical HZ is not well known because of the lack of a reliable geological surface history of Venus beyond about 1 billion years due to resurfacing of the stagnant lid, which allows for the possibility of a liquid surface ocean, however it does not stipulate a liquid ocean surface. Climate models also show limitations due to unknown cloud feedback for higher stellar irradiation.

Therefore we concentrate of the Earth equivalent orbital distance in this paper and use the empirical HZ limits as an example for an orbital distance range. The outer HZ limit agrees in the 3D and 1D model and are therefore not given in separate columns in Table 1. The orbital distance of the HZ boundaries can be calculated from $S_{eff}$ using equation 2:

$$d = \sqrt{\frac{L/L_{sun}}{S_{eff}}} \quad (2)$$

where $L/L_{sun}$ is the stellar luminosity in solar units and $d$ is the orbital distance in AU.

Models of the Habitable Zone take into account that cool red stars which have a stellar energy distribution that peaks at a redder wavelength than the Sun, heat the surface of an Earth-like planet more effectively. This effect is not included when calculating the 1AU equivalent distance, but it is included in atmospheric models, which generated the concept of the Habitable Zone. The stellar energy distribution depends on spectral type and changes with age. A star's radiation shifts to longer wavelengths with cooler surface temperatures, which makes the light of a cooler star more efficient at heating an Earth-like planet with a mostly $N_2$-$H_2O$-$CO_2$ atmosphere (see e.g. Kasting et al. 1993). This is partly due to the effectiveness of Rayleigh scattering, which decreases at longer wavelengths. A second effect is the increase in NIR absorption by $H_2O$ and $CO_2$ as the star's spectral peak shifts to these wavelengths, meaning that the same integrated stellar flux that hits the top of a planet's atmosphere from a cool red star warms a planet more efficiently than the same integrated flux from a star with a higher effective surface temperature. Stellar luminosity as well as the SED change with stellar spectral type and age, which influences the orbital distance at which an Earth-like planet can maintain climate conditions which allow for liquid water on its surface (see review by Kaltenegger 2017).

2.2 How the TESS Habitable Zone Star Catalog was created

We base our analysis on version 7 of the TESS Input Catalog (TIC-7), and the accompanying Candidate Target List (CTL-7.02), described in Stassun et al. (2018), and in the TIC Data Release Notes[1]. The TIC includes all stars in the sky down to the magnitude limits of all-sky photometric catalogs, most significantly 2MASS. Stars in the CTL are selected as high-quality targets for TESS observations, consisting primarily of bright, cool, dwarf stars. CTL-7 contains roughly 3.75 million stars, of which approximately 400,000 will ultimately be selected as TESS 2-minute observing targets for transit searches. The physical parameters of the stars in the CTL are compiled from a variety of catalogs, and also calculated using a set of empirical relations, all described in Stassun et al. (2018). Those physical parameters include the stellar temperatures, masses, radii, and luminosities.

---

[1] https://docs.google.com/document/d/1BQ4txZ5YmTMwNLiI5ilc-YHkYZqAMCg5HHFaTLoI33w/edit?usp=sharing

Since CTL-7.02 was assembled, the second data release from the Gaia mission (Gaia DR2) became available (Gaia Collaboration 2016, 2018). The release includes updates parallaxes for nearly all stars in the CTL. As discussed by (Bailer-Jones 2015), the simple $1/\pi$ distance estimator becomes sensitive to the assumed prior when the relative uncertainty in the parallax is larger than ≈20%. While the majority of the stars in our sample satisfy this criterion, for completeness we elected to adopt distances for all of the stars using an explicit Bayesian prior. Specifically, we use the distance estimates from Bailer-Jones et al (2018), which invokes a weak prior based on a simple Milky Way model for the stellar density distribution as a function of Galactocentric radius and height above the Galactic plane.

Following Sullivan et al. 2015, we first use a T-mag cutoff to limit our sample. Faint stars will be more difficult to detect transits of, obtain dynamical confirmation of any planets around them, or probe the atmospheres of such planets. To identify the T-mag cutoff, we selected a star with $T_{eff}$ = 3500K as our fiducial star, just as Sullivan selected a star with $T_{eff}$ = 3200K. Our choice differs from Sullivan because those authors did not have access to the actual TIC and CTL. We are able to identify $T_{eff}$ = 3500K as the most common $T_{eff}$ for a cool star in the CTL. At that $T_{eff}$, the stellar radius is $0.37R_{Sun}$, and a transit of an Earth-sized planet around a star that size would yield a transit depth of 600ppm. Following the same procedure as Sullivan, we then set the T-mag limit to correspond to the magnitude where the per-point photometric precision of TESS is equal to the fiducial transit depth, which for 600ppm is T = 12.

The TESS photometric bandpass is very broad (see Figure 1, Ricker et al 2016, Sullivan et al 2015). TESS will monitor a much larger sample of M stars compared to Kepler, thus the bandpass extends further to red wavelengths. It is broadly similar to a Cousins I-band filter. As shown in Fig.1, stars with different $T_{eff}$ can have similar T-mag. Thus our catalog provides targets for a wide-range of ground and space-based telescopes, which are optimized for different wavelengths.

A number of empirical relations have been defined (Stassun et al 2018) for conversion to T magnitude from many other standard filters and systems. To provide a very crude sense of the relationship between the TESS and Gaia magnitudes for the entire TIC, T = G - 0.5 with a scatter of < 1mag; that is, T is on average roughly one-half magnitude brighter than G.

We then calculate how long each of the CTL stars would be observed by TESS in the prime 2-year mission. We use a parameterized estimate (Barclay et al. 2018) based on the ecliptic latitude of the stars to calculate the average number of TESS sector observations each star would receive. For each star, we calculate the Earth equivalent orbital distance based on the star's luminosity and the corresponding orbital period. We then analyze whether the star would be observed for long enough by TESS for the detection of two transits for a planet at that orbital distance. Furthermore we calculate different limits of the HZ as described in Section 2.1 and redo our analysis for the concept of the empirical HZ.

2.3 REDETERMINATION OF RADII AND IDENTIFICATION OF RED GIANT CONTAMINANTS USING GAIA DR2 DATA

To identify evolved stars in our initial CTL-7 sample, we use the Gaia DR2 data to update the distance and radii assumed in CTL-7. 57 stars in our initial CTL-7 sample do not have reported Gaia DR2 parallaxes, and we were unable to match them to any Gaia source; these 57 stars in general appear to have very high proper motions (>100 mas/yr). Due to the unknown distance we remove these 57 stars from our analysis. We retain the effective temperatures from the current CTL-7 but apply the Gaia DR2 based distance estimator from Bailer-Jones et al. (2018) to rederive the stellar radii for our sample. This enables a more robust elimination of evolved stars from the sample. Current CTL-7 includes a likely contamination by red giants of about 2%, or about 30 stars in the sample used in this paper. We use the Gaia DR2 distances and recalculated the stellar radii from the Gaia DR2 G magnitudes, the Gaia bolometric correction ($BC_G$) computed from the effective temperature (according to Andrae et al. 2018), and the distance (Bailer-Jones et al. 2018).

The $BC_G$ relation is formally valid only for $T_{eff}$ > 3300K, therefore for stars whose temperatures had been estimated in the TIC-7 from the specially curated Cool Dwarf List (Stassun et al. 2018, Muirhead et al. 2018), we retained the stellar radius originally determined in the CTL; these are cool stars with high proper motions and therefore should be reliably nearby cool dwarfs.

Figure 2 shows our sample in the radius vs. temperature plane, where the radii have been updated from the CTL-7 as described above. The vast majority of the sample follows the expected relationship for MS cool dwarfs. However, a small number of stars have much larger radii; these are evolved stars that were incorrectly labeled as dwarfs in CTL-7. A cutoff of 1.5 $R_{Sun}$ effectively demarcates these evolved stars beyond the scatter in the MS population. The figure also highlights stars whose Gaia DR2 based distances are 300pc or beyond, which principally identifies the same evolved stars with radii larger than 1.5 $R_{Sun}$, but also identifies three additional stars that were putative cool dwarfs but are evidently at large distances such that they must be more luminous (i.e., larger) than expected. There are 35 stars in our sample that are identified as evolved stars by these criteria.

As a consistency check, we compare our derived stellar radii with those reported in the Gaia DR2 database for 813 of the stars in our sample (Gaia DR2 includes radius estimates for only about half of our sample). The agreement overall is excellent, except again for a small number of outliers already identified above. In addition, we note that the radii of the coolest stars are in general over-estimated by Gaia DR2, which is expected given the limitations of the Gaia DR2 stellar properties pipeline at temperatures below about 3500K (Andrae et al. 2018).

Figure 2 (right) shows how the radii newly determined in this paper compare to those from the original TIC. While planets orbiting evolved host stars might also host habitable planets (see e.g. Ramirez & Kaltenegger 2017), due to the short time a planet can spend in the RG HZ, we excluded RGs from our catalog.

In summary, we have verified that nearly all of our selected sample indeed consists of cool dwarf stars, based on radius estimates updated from the CTL using the newly available Gaia DR2 parallaxes. We are able to eliminate 35 stars as being more evolved (having larger radii) than had been estimated in the TIC-7; these 35 stars constitute about 2% of our sample, consistent with the TIC-7 contamination by red giants of 2% (Stassun et al. 2018). Here we identify and remove these evolved contaminants by requiring $R_{star} < 1.5\ R_{Sun}$ and d < 300 pc, leaving a final analysis sample of 1822 cool, nearby dwarfs.

2.4 WHICH PLANET SIZES CAN TESS DETECT DURING ONE TRANSIT?

To analyze which planets TESS can detect during one planetary transit, we first calculate the maximum transit time for each planet at the Earth-equivalent orbital distance around each of the stars in the TESS Habitable Zone Star Catalog. To estimate the noise, we adopted the approximate polynomial form of the TESS noise model from [Oelkers & Stassun (2018)](Oelkers & Stassun (2018)). This is a slightly updated version of the original TESS noise model from Sullivan et al. (2015), and uses the pre-launch systematic noise floor of 60 ppm hr^-0.5. Using the maximum transit time of each planet at the Earth equivalent orbital distance for their host star, we analyze whether TESS can detect a transiting planet of a certain size during one transit, with the minimum requirement being that the planet's signal is the same strength as the noise signal. We assume different sizes for the hypothetical planets between 1 and 2 Earth radii for our analysis. Note that number of stars around which TESS can detect transiting planets of a certain size is not the same as the number of planets we expect to find because the number of expected planet detections depend among different factors on the intrinsic occurrence rate of small planets in the habitable zones of these stars, the alignment of such planets orbiting stars in the sample we have presented, and the characteristics of the TESS data that would allow such transiting planets to be confidently detected, most with a small number of transits seen (see discussion 4.3).

3. RESULTS

3.1 FOR HOW MANY STARS CAN TESS FIND PLANETS AT EARTH-EQUIVALENT DISTANCE?

TESS can detect two transits of a planet orbiting at the 1AU equivalent distance for 1822 stars brighter than T=12 during its prime mission. The effective temperatures of these stars are roughly between about 2,700K and 5,400K, with distances up to about 300 parsec. The top panel of Fig. 3 shows the distribution of the TESS Habitable Zone stars on the sky, which a strong concentration around the ecliptic poles, where the TESS observational sectors overlap and allow for longer observation times. The size of the circles indicate the TESS magnitude with larger circles representing stars with higher apparent brightness.

The TESS Habitable Zone Stellar Catalog is populated by cool stars, which reflects the decrease of the Earth-equivalent orbital distance with decreasing luminosity of the star. The period of planets orbiting cool stars at these orbital distances are short enough so that 2 transits can be detected during the nominal TESS observation time of these stars. Due to TESS's observing strategy, TESS

Habitable Zone Catalog stars with higher effective surface temperature are located in the regions around the ecliptic pole, where the observing segments overlap and allow for longer observations. 137 of the stars of the TESS Habitable Zone Catalog are located in the JWST continuous viewing zone, which is located +/-5 degrees from the ecliptic poles. These stars have effective temperatures between 3,000K to 5,200K and distances up to about 230 parsec, indicated in Table 2 with the tag "JWST". Out to 20 degrees off the pole, JWST can observe a star for a minimum of 250 days out of a year. 1,350 stars of the TESS Habitable Zone Star catalog are located within 20 degrees from the ecliptic poles (see Fig. 3 for the subset of stars which are (3$^{rd}$ row) in the continuous viewing zone and (bottom) in the 250 day observation viewing zone of JWST).

Table 2 lists the stars sorted by decreasing apparent separation of the 1AU equivalent orbital distance as seen from Earth in milliarcseconds (mas). The column called "Sector" shows whether the star has already been observed with TESS in Sector 1 through 4. The catalog can be downloaded under https://filtergraph.com/tess_habitable_zone_catalog.

3.2 FOR HOW MANY STARS CAN TESS DETECT PLANETS IN THE WHOLE HABITABLE ZONE?

For 227 stars brighter than T=12, TESS can detect 2 transits of a planet through the whole empirical Habitable Zone, planets receiving stellar irradiation comparable to Recent Venus to Early Mars equivalent. Probing the whole Habitable Zone will give us a deeper understanding of the environment, formation and the distribution of planets in the Habitable Zone of stars. These 227 stars have effective temperatures between about 2,800K to 4,000K for distances up to about 100 parsec. The second row of Fig. 3 shows the distribution of these stars by distance based on Gaia DR2 data and the distribution of the stars in ecliptic latitude. Table 3 shows this specific subset of the stars in Table 2, the stars for which TESS can probe the whole HZ for transiting planets, sorted by decreasing apparent separation of the outer limit of the empirical HZ in mas, indicated in Table 2 with the tag "HZ".

3.3 FOR HOW MANY STARS CAN TESS DETECT A SMALL PLANET RECEIVING EARTH-LIKE IRRADIATION?

For 408 of the TESS HZ catalog stars, TESS has the sensitivity to detect a 1 Earth-size planet orbiting at the 1AU-equivalent distance, with a minimum requirement that the transiting planetary signal is bigger than the noise signal during one transit. For 1690 stars TESS can detect a planet with 1.6 Earth radii and for all 1822 stars TESS can detect a planet of 2 Earth radii or larger during one transit. The boundary between rocky and gaseous planets is not well understood yet, commonly values between 1.6 and 2 Earth radii are used as boundaries between rocky and gaseous planets (see e.g. Rogers 2015, Wolfgang et al 2016 and a discussion in Kaltenegger 2017). Figure 4 shows the transit signal for the subset of stars, where TESS can detect (a) a 1 Earth-size planet during one transit and (b) the full set of stars for which TESS can detect 2 Earth-size planet. Note that adding transit observation will increase the transit signal while requiring a specific signal to noise ratio will change the number of stars TESS can probe for transiting planets as well (see detailed discussion in 4.3).

3.4 PROSPECTS FOR FOLLOW-UP CHARACTERIZATION OF TESS PLANETS

Due to the wide range of effective surface temperatures of the stars in the TESS HZ catalog, which uses a Tmag cutoff, a wide range of telescopes can provide follow-up observations for different subsets of stars – e.g. instruments optimized for cool red stars would be optimized for the cool star in our catalog.

As discussed in detail in Sullivan et al (2015) and Ricker et al (2016), follow-up observations will allow characterization of TESS planets. Because of the short periods, even low mass planets – we have assumed 1 Earth mass planets for our calculations shown in Fig.4 - produce a radial-velocity semiamplitude K between 50 and 0.5 m s$^{-1}$ putting them within reach of current and upcoming spectrographs. Note that these values are a conservative estimate because they increase with planetary mass and a rocky planet models predict 10 Earth masses for a 2 Earth radii planet.

Ground-based facilities and upcoming space-based facilities such as CHEOPS (Fortier et al. 2014) could use photometry to look for transit timing variations to improve estimates of relative planetary mass. Asteroseismology using data from TESS or the upcoming PLATO mission (Rauer et al. 2014) could further constrain the stellar radii.

The prospects for follow-up atmospheric characterization with JWST have been detailed by several teams (e.g. Clampin et al 2009, Deming et al 2009, Kaltenegger & Sasselov 2010, Kempton et al. 2018). Note that the TESS HZ catalog is focused on TESS's sensitivity to planets between 1 and 2 Earth radii receiving similar insolation to present-day Earth. Thus the nominal calculations for planets using hydrogen atmospheres are not the best template. To detect features in an Earth analog transmission spectrum (see e.g. Kaltenegger & Traub 2009, Rauer et al. 2011, Kaltenegger 2017) will require a large telescope, and will be at the limit of technical capability of telescopes already being built like JWST (e.g. Deming et al. 2009, Kempton et al 2018) and ELT (e.g. Snellen et al 2013, Rodler & Lopez-Morales 2014). Such measurements are a baseline for future mission concepts like those proposed for the decadal survey for 2020.

Upcoming ground-based Extremely Large Telescopes (ELTs) are also designed to characterize extrasolar planets. The TESS planets will be excellent targets for observations due to the apparent brightness of their host stars. Comparing the angular separation (θ (arcsec) = a (AU) / d (pc); where a = planet semi-major axis, d = distance to star system) for the angular separation of the 1AU equivalent distance (Table 1) as well as the empirical Habitable Zone outer limit (Table 2) with the inner working angle (IWA) for a telescope, which describes the minimum angular separation at which a faint object can be detected around a bright star. This shows whether planets at such orbital distances could be remotely detected and resolved in the near future. For example the 38 m diameter Extremely Large Telescope should have an IWA of about 6 mas when observing in the visible region of the spectrum (assuming θ$_{IWA}$ ≈ 2(λ/D), where λ is the observing wavelength and D is the telescope diameter).

The apparent angular separation of the outer edge of the empirical HZ of the stars in the TESS Habitable Zone Star Catalog is between about 76 mas and 1.6 mas for the 227 stars, where TESS can probe the full HZ for transiting planets. The apparent angular separation of the 1AU equivalent orbital distance around all stars is between 33 and 0.3 mas. Assuming an IWA of 6 mas for the ELT, for 237 of the 1822 stars of the TESS Habitable Zone Stars Catalog, a planet at the 1AU equivalent distance could be resolved. Assuming an IWA of 6 mas, for 22 stars, a planet in the empirical HZ of the star could be resolved down to the inner edge of the HZ with ELT.

4. DISCUSSION

4.1 ASSUMED OBSERVING TIME WITH TESS

Because TESS started science observations in July 2018, the observing times for each star we examine are based on a parameterized estimate from (Barclay et al. 2018) based on the ecliptic latitude of the stars and the average number of TESS sector observations each star would receive by TESS in the 2-year prime mission lifetime. Note that this parameterization does not account for whether or not the observations will be continuous; for some host stars, e.g. stars observed in sector 1 and 13, the observations will cover 54 days, however, as the observing time will not have been undertaken continuously, a transit seen only once in one sector can not easily be combined with a transit seen only once in the second sector, and such a planet will need to be observed further to determine their orbit. The tables also show whether the star has already been observed in Sector 1 through 6 of the TESS mission.

4.2 ALTERNATIVE HABITABLE ZONE LIMITS

There is not one common irradiation limits used throughout the literature for the boundaries of the HZ. That stems from different teams using different irradiation limits for the boundaries of liquid water on the surface of a rocky planet. We focus on the 1AU equivalent flux in this paper and use the empirical HZ limits to explore the sample of stars where TESS can observe the full HZ. We use the inner limit being set by the irradiation Venus received when we know it did no longer hold liquid surface water (Recent Venus, see Kasting et al. 1993) assuming $N_2$-$CO_2$-$H_2O$ atmospheres. Note that the inner limit is not well known because of the lack of a reliable geological surface history of Venus beyond about 1 billion years due to resurfacing of the stagnant lid, which allows for the possibility of a liquid surface ocean, however it does not stipulate a liquid ocean surface. Using radiation higher than that limit for the inner edge of a Habitable Zone would have to explain why such planets could maintain their surface water in contrast to Venus (some small irradiation increase can be argued for dry desert planets, see Abe et al. 2011).

Using 3D Global Climate models, the inner edge of the Habitable Zone changes from the empirical value (see Table 1), however the outer edge of the HZ boundaries for $N_2$-$CO_2$-$H_2O$ atmospheres is similar between models. Thus the number of stars where TESS can probe the whole HZ does not change when using HZ limits derived from 3D models, because it is determined by the largest period that TESS is sensitive to a transiting planet at, which is set by the furthest distance from the star, which is the outer edge of the HZ. However the boundaries of the HZ and especially the outer edge of the HZ can change with atmospheric composition of the planet (e.g. by adding additional greenhouse gases like H, as discussed e.g. by Pierrehumbert & Gaidos 2011, Ramirez & Kaltenegger 2017) or $CH_4$ (Ramirez & Kaltenegger 2018). A lively discussion exists in the literature (see review by Kaltenegger 2017 and references therein) on the influence of rotation rate, cloud coverage and hazes on the limits of the HZ, using extrapolation of 3D models to different environments.

With no clear answer, we have used the empirical HZ in this paper to compare the number of stars where TESS can probe the whole HZ for transiting planets to the number of stars where TESS can detect planets receiving Earth-analog radiation. Note that being in the HZ does not necessarily mean that a planet is habitable, and in-depth follow-up spectral observations of their atmospheres are needed to characterize planets and search for signs of life (see review Kaltenegger 2017)

4.3 HABITABLE PLANET YIELD

While the catalog presented here provides a target list for investigations of the TESS stars for a habitable planet search, the actual yield of habitable planets remains unknown and is not the focus of this paper. The number of expected planet detections depend on a number of factors, including the intrinsic occurrence rate of small planets in the habitable zones of these stars, the fortuitous alignment of such planets orbiting stars in the sample we have presented, and the characteristics of the TESS data that would allow such transiting planets to be confidently detected, most with a small number of transits seen.

Several studies indicate that the occurrence rate of rocky planet in the HZ of cool stars are higher than for hotter stars (see e.g. Dressing & Charbonneau 2013, Petigura et al. 2013), however the numbers vary between studies (see e.g. Savranksy et al 2018 for a recent comparison. Note that these values have changed due to updates for distances by GAIA DR2 (see e.g. John et al 2018, Berger et al 2018). Several previous studies have estimated the transiting planet yield for TESS. Sullivan, et al (2015) predicts about 14, Bouma et al (2017) predicts 11, and Barclay et al (2018) predicts 9 detected planets with radii below 2 Earth radii in the HZ.

Our catalog does not attempt to revisit those calculations, which were based on average calculated stellar and planet distributions which did not include GAIA DR2 data. Rather, we provide the specific targets that will be worth special attention by TESS and the community in order to focus the search for habitable planets. However note, that the number of stars that can be probed for planets, is not the same as expected detected planets the so called planet yield.

As a very simplified example, assuming a 10% transit probability, a 25% occurance rate of planets out to 1AU equivalent orbital distance for all stars in our catalog would predict about 10 one-Earth-size planets, detected during one transit. Assuming a 50% occurance rate would predict 20 one-Earth-size planets out to the 1AU-equivalent distance. For 2 Earth-size radii the number increases to 45 planets out to the 1AU equivalent distance for an occurance rate of 25%, and 90 transiting planets for a 50% occurance rate, which TESS can detect during one. These numbers are only first order estimates to illustrate the point that the number of stars in the TESS Habitable Zone catalog does not equal the number of expected planets.

Note that TESS can probe the whole HZ for 227 stars brighter than T=12, however it can probe the HZ partially – out to Earth-equivalent irradiation - for all 1822 stars in the TESS Habitable Zone Star Catalog, thus an estimate for expected detection planets in the Habitable Zone would require a detailed model of planet distribution versus orbital distance, which is highly uncertain.

This paper identifies the host stars brighter than T=12, around which TESS

has the sensitivity to detect transiting planets receiving Earth-like irradiation as well as the subsample where TESS can observe the full Habitable Zone and thus can identify planets orbiting at these distances during its prime mission lifetime of 2 years. Observing the TESS Habitable Zone Catalog Stars will give us deeper insight into the occurrence rate of planets, out to Earth-analog irradiation as well as in the Habitable Zone, especially around cool stars.

4.4 COMPARISON TO OTHER CATALOGS IDENTIFYING NEARBY STARS FOR SEARCH FOR EXOPLANETS.

Several lists of the closest stellar targets for investigating habitable planets, with stars selected based on proximity, brightness, and other factors exist, e.g. Catalog of Nearby Habitable Systems (HabCat; Turnbull & Tartar 2003) lists over 17,000 nearby stars, the Target star catalogue for Darwin Nearby Stellar sample for a search for terrestrial planets (DASSC; Kaltenegger et al. 2010), which identified the best 1229 single main sequence stars for follow up observations, and the Catalog of Earth-Like Exoplanet Survey Targets (CELESTA; Chandler, McDonald, & Kane 2016), which lists 37,000 nearby stars. DASSC and CELESTA also provide habitable zone size estimates.

What distinguishes this catalog from previous work like HabCat, DASSC and CELESTA is that the stars included here are specifically selected because they are observed sufficient time by TESS to be able to detect 2 transits of planets which receive similar irradiation to Earth, and that TESS can detect planets down to 2 Earth-radii during one transit. All of these stars are included in the TESS exoplanet Candidate Target List, ensuring that they will also have 2-minute cadence observation (provided they do not fall in TESS camera pixel gaps), providing a specific catalog for the TESS mission of stars, where the 1AU equivalent distance can be probed for transiting planets.

5. CONCLUSIONS

We present the Transiting Exoplanet Survey Satellite Habitable Zone Stars Catalog, a list of 1822 nearby stars with TESS magnitude brighter than 12 with reliable Gaia DR2 distance. Around these stars TESS has the sensitivity to detect 2 transits of planets, which receive Earth analog irradiation during the 2 year prime mission. For all these stars TESS has the sensitivity to detect planets down to 2 Earth during one transit. For 408 of these stars, TESS can detect transiting planets down to 1 Earth-size during one transit. The catalog stars' effective temperatures are between about 2,700 K to 5,400 K. 137 of these stars are located in the JWST continuous viewing zone. 1350 of these stars are located in the 250 day JWST viewing zone, +/- 20 degrees off the ecliptic poles. For 227 stars in the TESS Habitable Zone Stars Catalog, TESS can probe the full extent of the Empirical Habitable Zone for transiting planets.

Observing the TESS Habitable Zone Catalog Stars will give us deeper insight into the occurrence rate of planets, out to Earth-analog irradiation as well as in the Habitable Zone, especially around cool stars.

Table 2 and 3 show the stars for which TESS can detect 2 transits for planets at 1AU equivalent distance and the subset of stars where TESS can observe planets throughout the whole Habitable Zone, respectively. The tables sort the stars in the TESS Habitable Zone Star Catalog by apparent angular separation, showing that the 1AU equivalent distance for hundreds of these stars can be resolved for upcoming Extremely Large

Telescopes, making them excellent targets for atmospheric characterization in the near future.

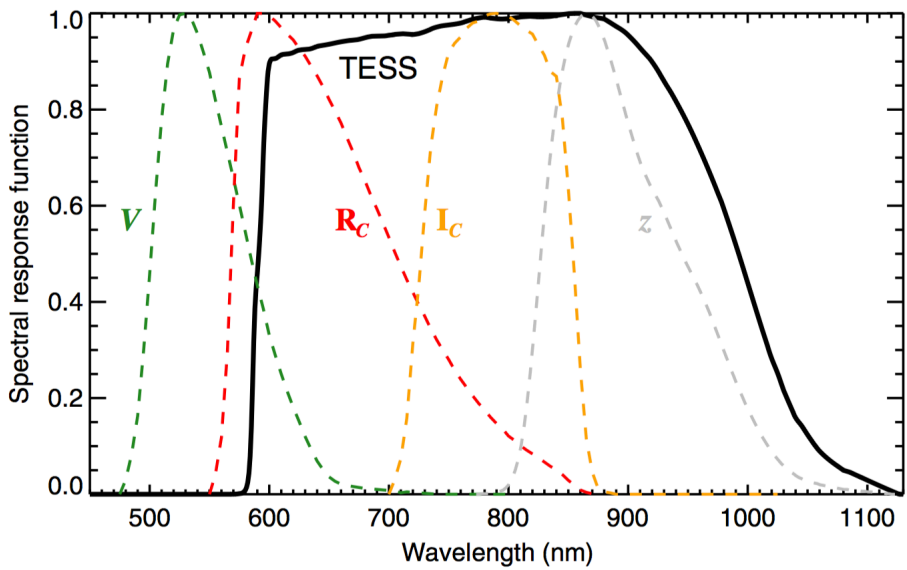

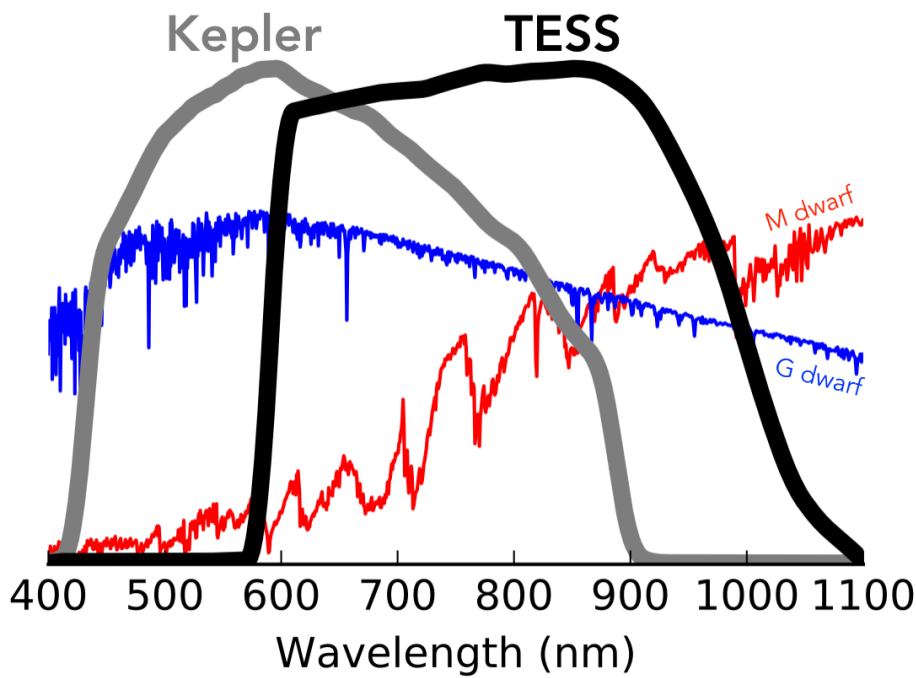

**Figure 1: (top)** TESS spectral response function (black line), defined as the product of the long-pass filter transmission curve and the detector quantum efficiency curve. Johnson-Cousins V, $R_C$, and $I_C$ filter curves and the Sloan Digital Sky Survey z filter curve. Each of the functions has been scaled to have a maximum value of unity. Image Credit: Ricker et al. (2016) **(bottom)** TESS bandpass compared to Kepler (Image Credit: Zach Berta-Thompson with data from Sullivan at al. 2015)

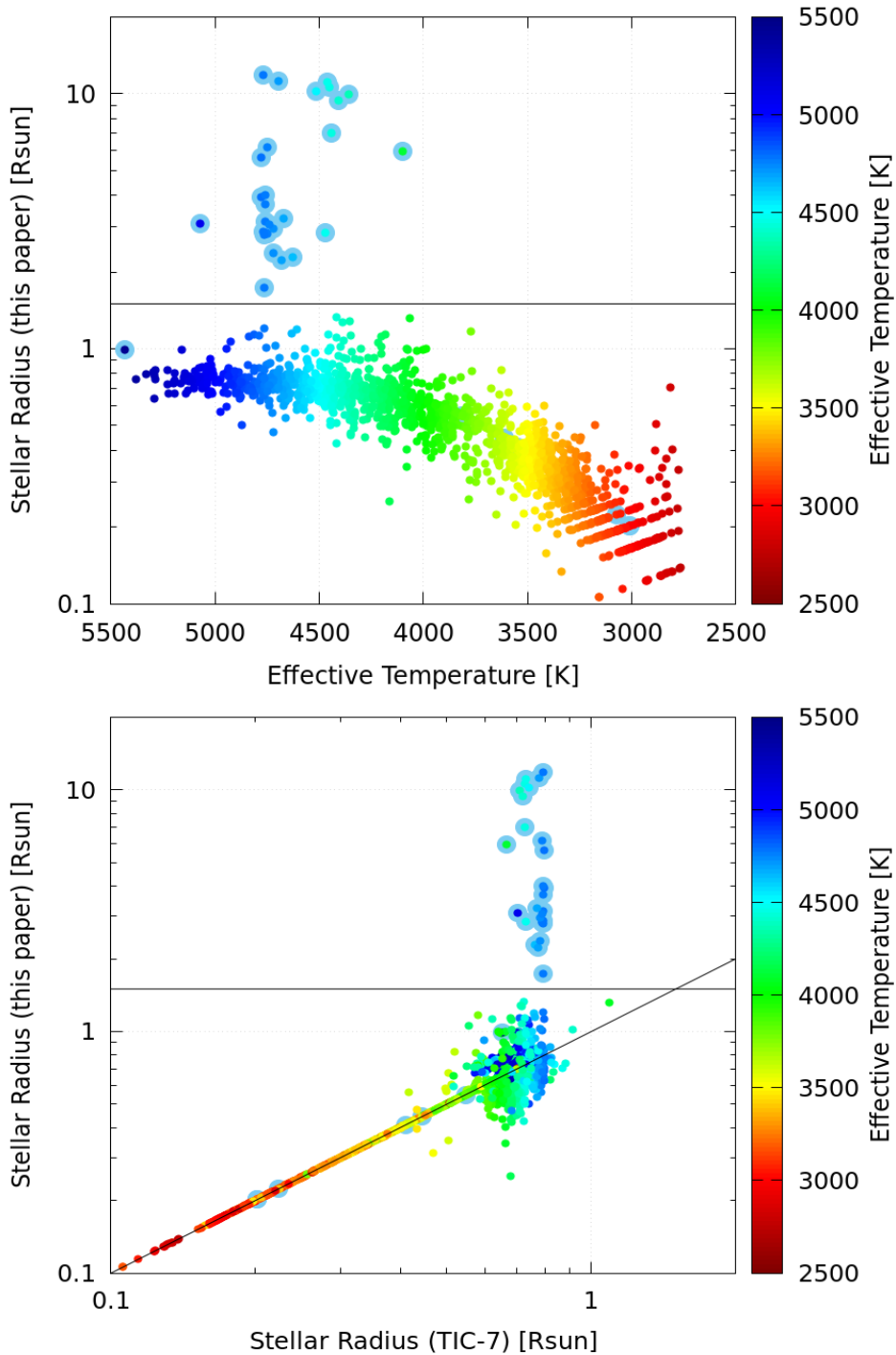

**Figure 2:** Stellar radius versus (top) effective temperature for our TIC-7 study sample. The effective temperatures are taken from the CTL but the radii are updated from the CTL using the newly available Gaia DR2 parallaxes. Highlighted in blue are evidently evolved stars that we identify and eliminate as $R_{star} < 1.5\ R_{Sun}$ or $d > 300pc$. (bottom) The stellar radii in this paper are compared against the stellar radii as originally reported in TIC-7. The color bar represents the effective stellar temperature.

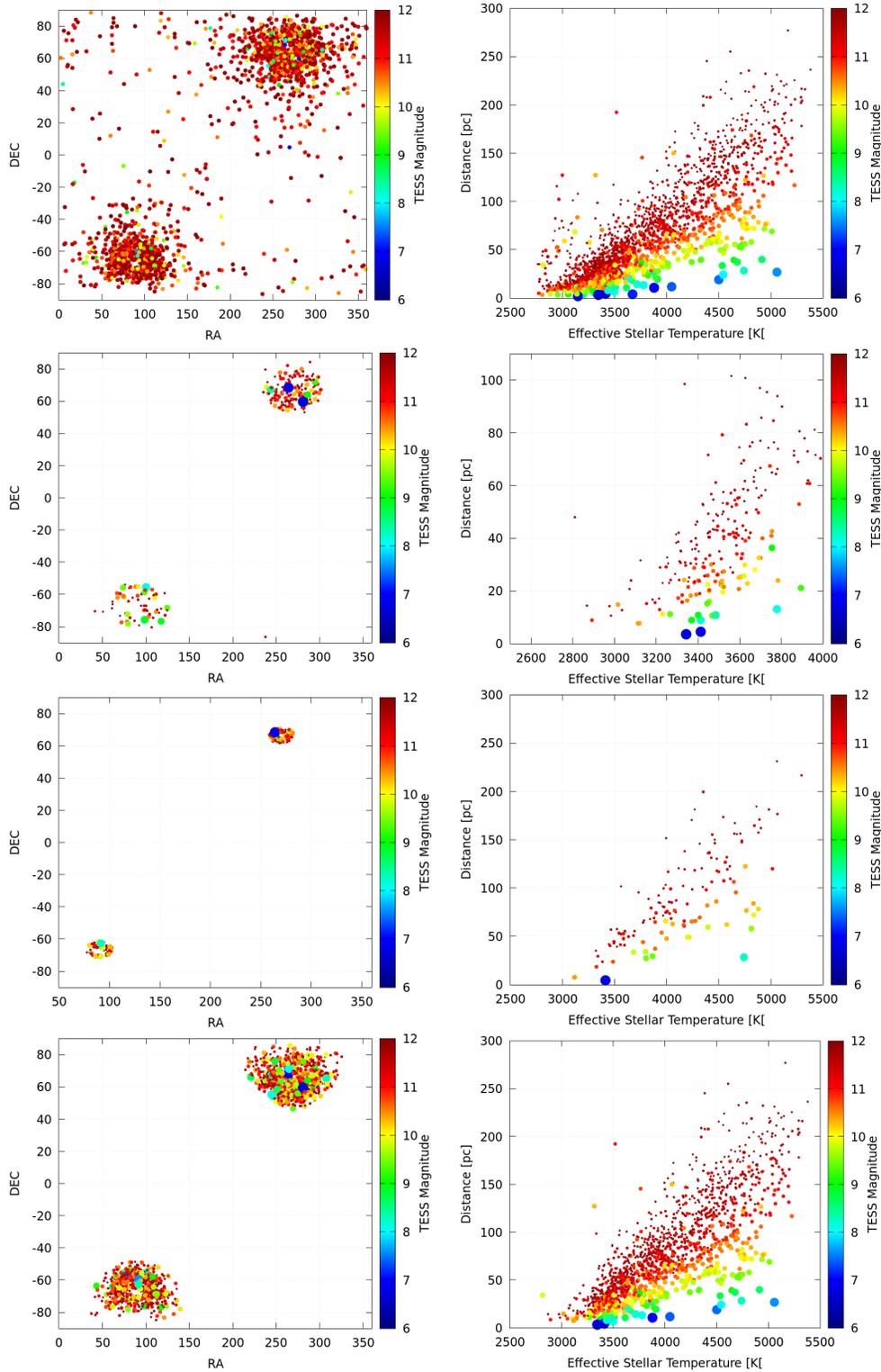

**Figure 3: (left) Distribution and (right) Distance of these stars from the Sun versus Effective stellar temperature of the stars on the sky where TESS can detect planets (top) out to the 1AU equivalent distance and (2$^{nd}$ row) throughout the whole empirical HZ (3$^{rd}$ row) JWST continuous viewing area and (bottom) a 250day JWST viewing zone.**

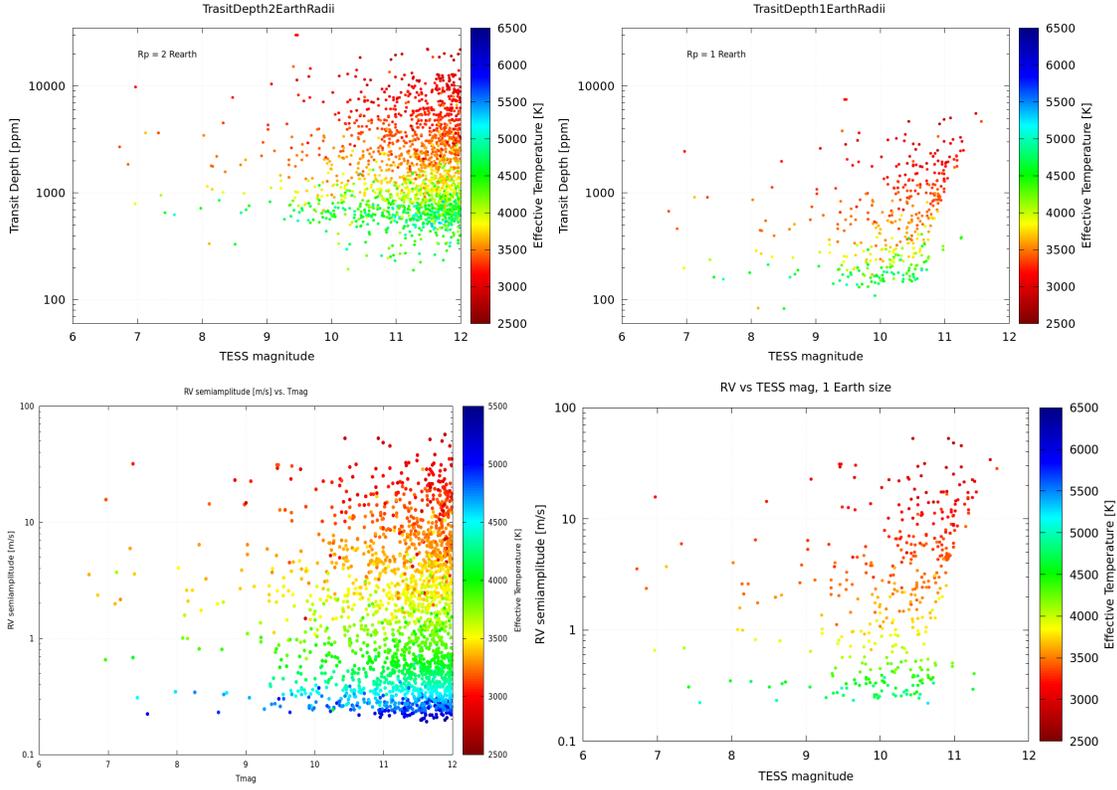

**Figure 4:** Modeled Transit depth (top) for each star of the TESS Habitable Zone Star Catalogue, where TESS can detect a transiting planet of (left) a 2 Earth radii and (right) a 1 Earth radius size during one transit. Modeled Radial-velocity semiamplitude K in m s$^{-1}$ (bottom) assuming a 1 Earth mass planet orbiting at a distance where it receives the same radiation as Earth does from the Sun, from its host star.

**Table 1:** Constants to compute the empirical boundaries of the Habitable Zone using eqn. 1 (Kopparapu et al.2013, Ramirez & Kaltenegger 2016).

| Constants | Recent Venus limit Inner edge | 3D model limit Inner edge | Early Mars limit Outer edge |
|---|---|---|---|
| $S_{sun}$ | 1.7665 | 1.1066 | 0.324 |
| A | 1.3351e-4 | 1.2181e-4 | 5.3221e-5 |
| B | 3.1515e-9 | 1.534e-8 | 1.4288e-9 |
| C | -3.3488e-12 | -1.5018e-12 | -1.1049e-12 |

**Table 2: Stars where TESS can detect planets out to the 1AU equivalent distance, sorted by decreasing apparent angular separation in milliarcsec (mas), the full table is available online [a].**

**Table 3: Stars where TESS can detect planets throughout the empirical HZ, sorted by decreasing apparent angular separation in milliarcsec (subset of Table 2), full table is available online[a].**

**a) Legend for Table 2 and 3**
Units  Label   Explanations
--------------------------------------------------------------------------------
---      TIC     TESS Input Catalog identifier
K        Teff    Effective temperature
mag      Tmag    TESS broad band magnitude
pc       Dis     Distance
d        Obs     Observing time
AU       aRV     Recent Venus orbital separation
AU       aEA     Earth Analog orbital separation
AU       aEM     Early Mars orbital separation
d        PerRV   Recent Venus orbital period
d        PerEA   Earth Analog orbital period
d        PerEM   Early Mars orbital period
deg      ELAT    Ecliptic latitude
deg      GLAT    Galactic latitude
deg      RAdeg   Right Ascension in decimal degrees (J2000)
deg      DEdeg   Declination in decimal degrees (J2000)
mas      asepRV  Recent Venus apparent angular separation
mas      asepEA  Earth Analog apparent angular separation
mas      asepEM  Early Mars apparent angular separation
h        Ttime   Transit time
---      Gaia    Gaia catalog identifier
---      2MASS   2MASS all sky survey catalog identifier
---      HZ      TESS can probe the full habitable zone for transiting planets (1)
---      JWST    Star is in the JWST continuous viewing zone (1)
---      Earth   TESS can detect an Earth-zed planet during one transit (1)
---      S1      Observed in TESS mission Sector 1 (1)
---      S2      Observed in TESS mission Sector 2 (1)
---      S3      Observed in TESS mission Sector 3 (1)
---      S4      Observed in TESS mission Sector 4 (1)
---      S5      Observed in TESS mission Sector 5 (1)
---      S6      Observed in TESS mission Sector 6 (1)
--------------------------------------------------------------------------------
Note (1):
   1 = yes;
   0 = no.
--------------------------------------------------------------------------------

data available online: https://filtergraph.com/tess_habitable_zone_catalog